\documentclass[oldversion]{aa}
\pdfoutput=1

\usepackage[intlimits]{amsmath}
\usepackage{txfonts}
\usepackage[american]{babel}
\usepackage{graphicx}
\usepackage{xspace}
\usepackage{url}

\usepackage{natbib}
\newcommand{\Alfven}{Alfv\'{e}n\xspace}
\newcommand{\Rb}{\ensuremath{R_\bnd}}
\newcommand{\rb}{\ensuremath{r_\bnd}}
\newcommand{\bnd}{\ensuremath{\mathrm{b}}}

\newcommand{\simm}{\ensuremath{\mathord{\sim}}}

\newcommand{\abs}[1]{\ensuremath{\left|#1\right|}}
\newcommand{\vA}{\ensuremath{v_\mathrm{A}}}
\newcommand{\vfm}{\ensuremath{v_\mathrm{fm}}}

\newcommand{\vphimax}{\ensuremath{v_{\varphi,\bnd}^\text{max}}}
\newcommand{\betab}{\ensuremath{\beta_\bnd}}
\newcommand{\Mb}{\ensuremath{\mathsf{M}_\bnd}}
\newcommand{\Bb}{\ensuremath{B_\bnd}}

\newcommand{\cs}{\ensuremath{c_\mathrm{s}}}
\newcommand{\csb}{\ensuremath{c_\mathrm{s,b}}}
\newcommand{\cso}{\ensuremath{c_{\mathrm{s}0}}}

\newcommand{\evarphi}{\ensuremath{\hat{\vec{e}}_\varphi}}

\newcommand{\ex}{\ensuremath{\hat{\vec{e}}_x}}
\newcommand{\de}{\ensuremath{\mathrm{d}}}
\newcommand{\const}{\ensuremath{\text{const}}}
\newcommand{\degree}{\ensuremath{^\circ}}

\newcommand{\efrate}{\ensuremath{\mathcal{E}}}

\begin{document}

\title{Large jets from small-scale magnetic fields}
\author{R.~Moll}
\institute{Max-Planck-Institut f\"{u}r Astrophysik, Karl-Schwarzschild-Str. 1, 85748 Garching, Germany}
\date{Accepted on November 28, 2009}
\abstract{We consider the conditions under which a rotating magnetic object can
produce a magnetically powered outflow in an initially unmagnetized medium
stratified under gravity. 3D MHD simulations are presented in which the
footpoints of localized, arcade-shaped magnetic fields are put into rotation.
It is shown how the effectiveness in producing a collimated magnetically
powered outflow depends on the rotation rate, the strength and the geometry of
the field.  The flows produced by uniformly rotating, non-axisymmetric fields
are found to consist mainly of buoyant plumes heated by dissipation of
rotational energy.  Collimated magnetically powered flows are formed if the
field and the rotating surface are arranged such that a toroidal magnetic field
is produced.  This requires a differential rotation of the arcades' footpoints.
Such jets are well-collimated; we follow their propagation through the
stratified atmosphere over 100 times the source size.  The magnetic field is
tightly wound and its propagation is dominated by the development of
non-axisymmetric instabilities.  We observe a Poynting flux conversion
efficiency of over 75\% in the longest simulations. Applications to the
collapsar model and protostellar jets are discussed.}

\maketitle

\section{Introduction}

Models and simulations of jets produced by rotating magnetic fields generally
assume the existence of an ordered, axially symmetric large-scale field of
uniform polarity anchored in the central engine, starting with the original
models of jets from accretion disks by \citet{1976Bisnovatyi} and
\citet{1976Blandford}, or the magnetic supernova model of \citet{1970LeBlanc}.
While such ordered fields are the most effective in producing jets, the
question whether they actually exist in accretion disks or the core of a star
is still quite open.  The largest scale at which magnetorotational (MRI)
turbulence in accretion disks shapes the magnetic field structure is set by the
disk thickness, which in turn is much smaller than jets.  Collapsar cores are
also small compared to the expected GRB jet, and it is not at all clear why the
magnetic fields there should be ordered and axisymmetric. Large-scale fields
are not easily trapped by an accretion disk \citep{1989Ballegooijen}.  Without
a large-scale field protruding from the disk, one may still hope to launch
outflows by twisting fields inside the disk or by magnetic loops that extend
into the disk corona as e.g. in \citet{1979Galeev,1996Tout,2008Uzdensky}.

The huge range of involved length scales is a major issue in jet modeling, as
numerical simulations that cover all scales are not feasible to date.
Protostellar jets may be several parsecs long, launched by disks with sizes of
$\simm 100\,\mathrm{AU}$, i.e.  about a factor $10^4$ smaller
\citep[e.g.][]{2001Shepherd}.  Assuming that the ``launching scale'' is much
smaller than the disk, perhaps on the order of $\simm 1\,\mathrm{AU}$, one
obtains an even bigger contrast in length scales.  Supposing that jets in AGN
are launched at a few Schwarzschild radii from the central black hole and
taking Cygnus~A as an example, one obtains a ratio of $10^6$ between the jet
length and the size of the engine \citep{1998Krichbaum,2003Tadhunter}.  The
final jet properties are determined before it becomes ballistic, probably at
scales which are somewhat smaller than that of the largest visible structures,
but still considerably larger than the central engine.

There is some numerical evidence that small-scale fields can also be used to
generate jets.  Most of these simulations are promising in terms of outflow
production but limited to the immediate surroundings of the outflow-forming
disk.  Axisymmetric simulations of outflows generated with small magnetic loops
were done by \citet{1998Romanova,1999Turner,2002Kudoh}.  In their 3D
simulations of accretion flows, \citet{2004Kato} demonstrated that an initially
poloidal magnetic field confined within a rotating torus surrounding the
accreting black hole can give rise to a transient outflow driven by accumulated
toroidal fields in the form of a ``magnetic tower'' \citep{2003Lynden}.
\citet{2005DeVilliers} showed in simulations that loops of poloidal field in an
accreting torus may give rise to a large-scale poloidal field as the field
lines are stretched out in an axial outflow.  The inflation and disruption of a
magnetic loop outside a disk, caused by the generation of a toroidal field
through differential rotation, was observed by \citet{2009Fendt} in simulations
of outflows from star-disk magnetospheres.  The generation of magnetic flows
within unmagnetized surroundings has also been studied in laboratory
experiments \citep{2005Hsu,2009Ciardi}; such jets are strongly affected by
current-driven instabilities.

The aim of the calculations presented here is to see whether small-scale fields
in the form of loops anchored in a rotating disk can be used to produce jets of
significant length (compared to the size of the source).  In the cases studied
the flows propagate and are confined in an external unmagnetized atmosphere (as
opposed to \citealt{2008Moll} and \citealt{2009Moll}, hereafter Papers I\&II,
where we studied jets embedded in a large scale magnetically dominated
environment).  While still idealized, this addresses environments like
protostellar jets launched into a dense cloud, or GRB jets launched by a
collapsar core. To be investigated here are the circumstances under which
jetlike flows are formed that penetrate through the atmosphere instead of
dissipating in it.  As this is also a question of (non-axisymmetric) stability,
three-dimensional simulations are necessary.  One of the questions to be
answered is whether models of jets from small-scale magnetic fields are a
viable alternative to those based on the twist of large-scale fields.

\section{Models}
\label{sec:models}

\begin{figure}[t]
\includegraphics[width=\linewidth]{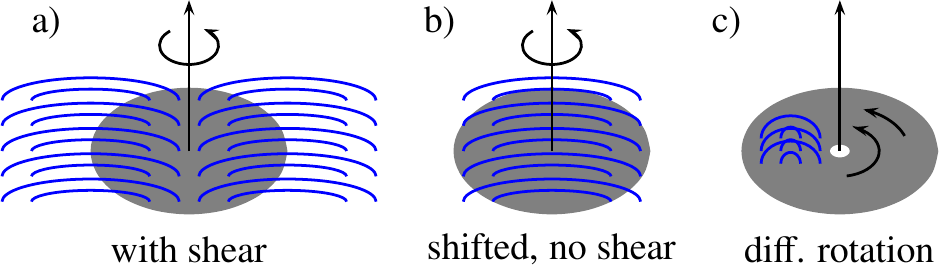}
\caption{Considered magnetic field geometries.  In case (a), the field lines
are one-sidedly anchored in a rigidly rotating disk. In case (b), the field
lines' footpoints are either both anchored in the disk, or they are both
anchored outside of it. In case (c), the field emerges from a limited area
inside the disk which is differentially rotating.}
\label{fig:emergecases}
\end{figure}

The primary ingredients in all our models are a rotating disk, implemented as a
boundary condition, and magnetic field loops anchored in and sticking out of
the disk.  The magnetic field geometries considered are sketched in
Fig.~\ref{fig:emergecases}.  In case (a), some of the loops have one footpoint
inside the disk while the other is anchored outside.  The resulting shear
motion of the footpoints generates a toroidal magnetic field.  In case (b), the
loops are arranged such that they are not sheared. Here, a toroidal field can
only be produced by the inertia of the material above the disk, such as in
conventional models of jets from large-scale magnetic fields.  This case would
be a model for a (solidly) rotating stellar core inside a nonmagnetic envelope.
Finally, we consider a case in which all loops are anchored in the disk and
shear is created through differential rotation (c). This case would be more
representative of small-scale fields generated in an accretion disk.

The magnetic field loops are established either by a suitable potential field,
imposed as initial condition (setup D), or by continuous ``injection'' of the
field from below the disk (setup E), see Fig.~\ref{fig:emergeschematic} and
Sect.~\ref{sec:setupE} for details.  While the magnetic field geometry is
similar in both cases, setup E (for ``Emerging field'') is meant as an
idealization of the field loops emerging from magnetic turbulence in an
accretion disk.

The model chosen as initial condition for the atmosphere is a hydrostatic
equilibrium stratification in the gravitational potential of a point mass (the
origin).  Temperature $T$ and density $\rho$ vary with distance $r$ from the
point mass as $r^{-1}$ and $r^{-3}$, respectively. The temperature is thus a
constant fraction of the virial temperature, as in advection-dominated
accretion flows. The density profile has the property that the amount of mass
in a cone centered on the origin is constant per decade in distance $r$.  The
stratification is thus scale-free, and a prospective jet encounters the same
amount of atmosphere mass per decade traveled.  With the equation of state used
in the calculations, an ideal gas with ratio of specific heats $\gamma=5/3$,
the stratification is convectively stable.

\section{Methods}
\label{sec:methods}

\subsection{Numerical MHD solver, grid and coordinates}
\label{sec:equations}

\begin{figure}[t]
\includegraphics[width=\linewidth]{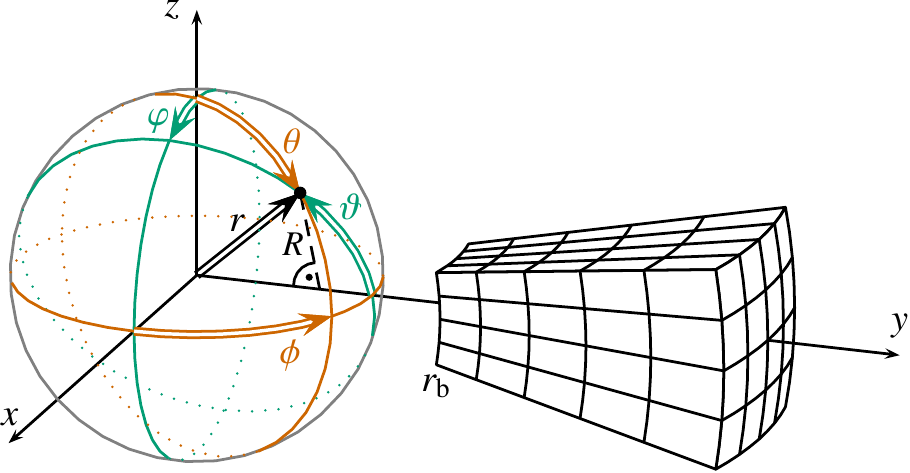}
\caption{Sketch of the coordinate system and the computational grid.  The jets
propagate in equatorial direction of the grid coordinates $(r,\theta,\phi)$.
To describe the results, the alternate spherical coordinate system
$(r,\vartheta,\varphi)$ is used, which takes the jet's central axis as the
polar axis.}
\label{fig:coords}
\end{figure}

We numerically solved the ideal adiabatic MHD equations with a static external
gravitational potential $\Phi \propto r^{-1}$ on a spherical grid
$(r,\theta,\phi)$.  The jets propagate in equatorial direction along the
$y$-axis ($\theta=\phi=\pi/2$), about which the computational volume covers a
range $\Delta\theta=\Delta\phi$ in the angular directions, see
Fig.~\ref{fig:coords}.  The spacing is uniform in $\theta$ and $\phi$ and
logarithmic in $r$.  Such a grid is more economical than a Cartesian one since
it expands in jet direction. Unlike a spherical grid in polar direction $z$, it
is free of singularities which may cause numerical problems or artifacts.

As jet physics is better described in a coordinate system which takes the jet's
axis $y$ as the polar axis, we introduce another spherical coordinate system
$(r,\vartheta,\varphi)$ for that purpose, see also Fig.~\ref{fig:coords}.  $R
\coloneqq r\sin\vartheta$ denotes the orthogonal distance to that axis
(cylindrical radius).

The simulations were performed with a newly developed Eulerian MHD code
\citep{2008Obergaulinger}.  It is based on a flux-conservative finite-volume
formulation of the MHD equations and the constraint transport scheme to
maintain a divergence-free magnetic field \citep{1988Evans}.  Using
high-resolution shock capturing methods \citep[e.g.,][]{1992LeVeque}, it
allows a choice of various optional high-order reconstruction algorithms and approximate
Riemann solvers based on the multi-stage method \citep{2006Toro}.  The
simulations presented here were performed with a fifth order
monotonicity-preserving reconstruction scheme \citep{1997Suresh}, together with
the HLL Riemann solver \citep{1983Harten} and third order Runge-Kutta time
stepping.

As is done in other astrophysical MHD simulations, the implicit assumption has
been made that the magnetic reconnection in the settings studied is of the
"fast" type, for which the reconnection implicit in the numerical diffusion of
the code may be a fair representation of the subgrid dynamics.

\begin{figure}[t]
\centering
\includegraphics[width=.9\linewidth]{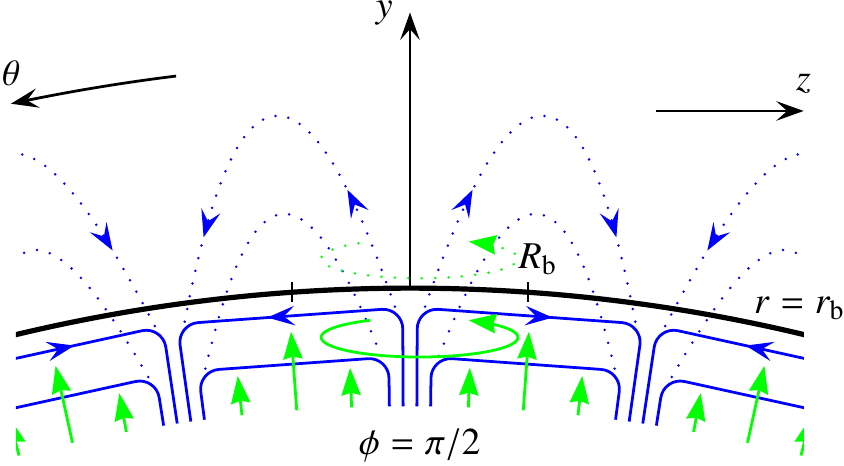}
\caption{Schematic of simulation setup E described in Sect.~\ref{sec:setupE},
in the variant sketched in Fig.~\ref{fig:emergecases}a.  A vertical velocity
field (green arrows) pushes the magnetic field (blue lines) from the lower
boundary ($r=\rb$) into the computational volume.  An azimuthal velocity field
within $R \le \Rb$ twists the created magnetic arcades (dotted blue lines)
which are one-sidedly anchored in the rotating region.}
\label{fig:emergeschematic}
\end{figure}

\subsubsection{Setup with an arcade-shaped, exponentially decreasing initial field (D)}
\label{sec:setupD}

The initial condition for the magnetic field in this case is a two-dimensional
potential field which is independent of the $x$-coordinate (see
Figs.~\ref{fig:coords},\ref{fig:emergeschematic} for the coordinate system) and
falls off exponentially with height $y$.  It is generated by the vector
potential
\begin{equation}
    \vec{A} = B_\bnd \lambda \sin\left(\frac{z}{\lambda}\right) \exp\left( \frac{\rb-y}{\lambda} \right) \ex,
\end{equation}
where $\lambda\pi$ is the width of the arcades and $\lambda$ is the scale
height of $B=\abs{\vec{B}}$. $B_\bnd$ is the field strength at the intersection of the
lower boundary $r=\rb$ with the central axis ($y$). To avoid numerical problems
at the lateral boundaries, we limit the number of arcs to 4, setting
$\vec{A}=\vec{B}=0$ for $\abs{z} > 2\lambda\pi$ and for $\abs{x} >
2\lambda\pi$.

The magnetic field is embedded in a spherically stratified atmosphere with $p
\propto r^{-4}$ and $\rho \propto r^{-3}$, held in hydrostatic equilibrium by
the static gravitational field $\Phi \propto r^{-1}$ of a point mass at the
coordinate origin.  Temperature and sound speed vary as $T \propto r^{-1}$ and
$\cs \propto r^{-1/2}$, respectively.  In the outer, unmagnetized region
($\abs{x},\abs{z} > 2\lambda\pi$), we compensated for the absence of magnetic
pressure by increasing the gas pressure. To maintain hydrostatic equilibrium,
the density is also increased correspondingly.

At the lower ($r=\rb$) boundary we maintain, through ``ghost cells'' outside of
the computational domain, an azimuthal velocity field $\vec{v}=v_\varphi
\evarphi$ corresponding to rigid rotation: $v_\varphi = \vphimax R/\Rb$ for $R
\le \Rb$ and 0 elsewhere.  All quantities except for $\vec{B}$ are fixed at
their initial values in the ghost cells;  $\vec{B}$ is extrapolated from the
interior of the domain. At the sides ($\theta$ and $\phi$) and top (upper $r$)
of the domain, we use open boundary conditions which allow for an almost
force-free outflow of material and cause no evident artifacts in the form of
reflections.

$\lambda$ is chosen such that in the $x=0$ plane $B_y$ is positive for $\abs{z}
< \Rb$ and negative for $\Rb < \abs{z} < 2\Rb$, i.e.  arcades which start at
$\abs{z} < \Rb$ from the $r=\rb$ surface have their second footpoint outside
the rotating region. This is achieved by choosing the length scale of the
arcade as $\lambda = 2\Rb / \pi$.

\subsubsection{Setup with magnetic field arcades emerging from the bottom boundary (E)}
\label{sec:setupE}

In this case, the initial condition is again an equilibrium stratification with
$p \propto r^{-4}$, $\rho \propto r^{-3}$ and $\Phi \propto r^{-1}$.  However,
unlike in setup D, the atmosphere is completely unmagnetized.

The magnetic field enters the domain through the lower boundary $r=\rb$, the
conditions of which are determined through ghost cells at $r<\rb$.  There, we
impose a transverse magnetic field $B_\theta$ ($\approx B_z$) with constant
amplitude $B_\bnd$ and a polarity that alternates with $z=\rb \cos\theta$ in
step sizes of $2\Rb$, the diameter of the rotating surface (described below).
Where the polarity of $B_\theta$ changes, the solenoidality of the magnetic
field is maintained by an appropriate $B_r$.  The field lines thus have the
shape sketched in Fig.~\ref{fig:emergeschematic}.  An approximate equilibrium
is maintained through lowering the gas pressure by the value of the magnetic
pressure, as far as this is possible ($p$ must not be negative).  For this to
work, $\betab \coloneqq 8\pi p_\bnd / B_\bnd^2$ must be greater than one, which
limits the possible field strengths to $B_\bnd < 1$ in our system of units
(described below).

To model the emergence of magnetic fields into the atmosphere we impose the
radial velocity field
\begin{equation}
    v_r = 0.9 \csb \abs{\sin \frac{z\pi}{2\Rb}}
\label{}
\end{equation}
in the ghost cells of the lower boundary.  The injection velocity is too small
to form a jet by itself: the maximum amounts to $41\%$ of the escape velocity
at the boundary, gas with this speed gets theoretically as far as $r \approx
1.2 \rb$ without acceleration other than gravity.  In addition to the radial
velocity field, we maintain an azimuthal velocity field $v_\varphi \propto R$
(rigid rotation) within $R\le\Rb$. The shape of the emerging magnetic field is
sketched in Fig.~\ref{fig:emergecases}a.  For the other boundaries, we use the
same outflow conditions as in the model described in the preceding section.

Two variations of the above-described setup have also been studied,
corresponding to the cases (b) and (c) described in Sect.~\ref{sec:models}.  In
the first, the positions where $B_\theta$ changes its polarity, and with it the
radial velocity field that injects the magnetic field, are shifted by $R_\bnd$
in $-z$ direction. Thus, both footpoints of an individual field loop rotate
with the same angular velocity.  In the second case (c), the magnetic field and
vertical velocity in the lower boundary are changed such that only a small,
confined arcade in $0.2 < R/R_\bnd < 0.8$ with a width of $0.2\Rb$ emerges, and
the azimuthal velocity field has a Keplerian profile with $v_\varphi = \vphimax
\sqrt{0.1\Rb/R}$ for $0.1 \le R/R_\bnd \le 1$. The inner edge ($R=0.2\Rb$) of
the arcade rotates twice as fast as the outer edge ($R=0.8\Rb$), the difference
in rotation velocity is $\vphimax/\sqrt{8}$.

\subsection{Parameters and units}
\label{sec:units}

\begin{table}
\caption{Normalization units}
\centering
\begin{tabular}{ccc}
\hline\hline
Quantity & Symbol(s) & Unit \\
\hline
    length          & $x$,$y$,$z$,$r$,$R$   & $l_0$ \\
    gas pressure    & $p$                   & $p_0$ \\
    density         & $\rho$                & $\rho_0$ \\
    velocity        & $v$                   & $\cso = \sqrt{ \gamma p_0 / \rho_0 }$ \\
    time            & $t$                   & $t_0 = l_0 / \cso$ \\
    energy flow rate& $\efrate$             & $p_0 l_0^3 / t_0$ \\
    magnetic flux density       & $B$       & $B_0 = \sqrt{8 \pi p_0} $ \\
\hline
\end{tabular}
\label{tab:units}
\end{table}

The models described above contain the following 6 parameters, not all of which
are independent: $r_\bnd$, $\Rb$, $\rho_\bnd$, $p_\bnd$, $B_\bnd$ and
$\vphimax$.  We eliminate the dependences by taking $l_0 \equiv 2\Rb$, $\rho_0
\equiv \rho_\bnd$ and $p_0 \equiv p_\bnd$ as units of length, density and
pressure, and expressing all physical quantities in terms of these, see
Table~\ref{tab:units}.  The remaining 3 parameters can then be expressed as
dimensionless numbers.  The first of these is $l_0/r_\bnd$, which is a measure
for the curvature of the rotating surface.  In all simulations presented here,
this curvature is small and probably does not influence the results
significantly.  The parameter has technical significance, however, because it
also controls the opening of the lateral boundaries.  The remaining two numbers
are chosen to be $\betab \coloneqq 8\pi p_0 / B_\bnd^2$, which controls the
strength of the magnetic field, and the Mach number $\Mb \coloneqq \vphimax /
\cso$, which determines the speed of rotation.  Note that in the emerging field
model (E), $\betab$ is a measure of the field strength but not a local
plasma-beta value, since the pressure entering its definition is not measured
at the same location as the field strength.  For the sake of clarity, we
usually omit the units in the presentation of the results. The concerned
quantity is then measured in terms of the associated normalization unit listed
in Table~\ref{tab:units}.

\section{Results}
\label{sec:results}

As explained in Sect.~\ref{sec:models} above, the calculations were done with
different models for the rotating magnetic field configuration.  It turns out
that some of these cases produce long-lived jets, others only transient flows
or flows that dissipate in the atmosphere close to the source.  The results
presented in the following subsection were obtained with setup D
(Sect.~\ref{sec:setupD}), those in the subsequent subsections were obtained
with setup E (Sect.~\ref{sec:setupE}).

\subsection{Transient jets with arcade-shaped initial field}

\begin{figure}[t]
\includegraphics[width=\linewidth]{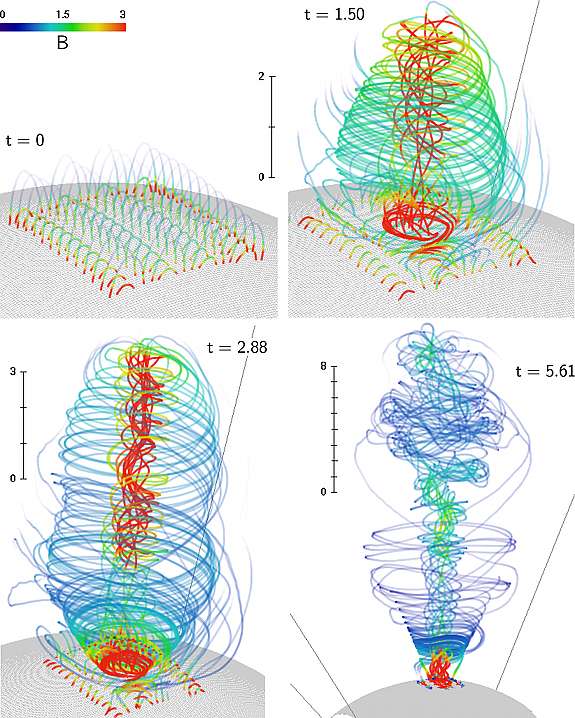}
\caption{Selected magnetic field lines in a simulation with an arcade-shaped
initial field (setup D). The lines are closing with the lower boundary,
ascending inside the jet and descending at the outside.}
\label{fig:flinesD}
\end{figure}

\begin{figure}[t]
\includegraphics[width=\linewidth]{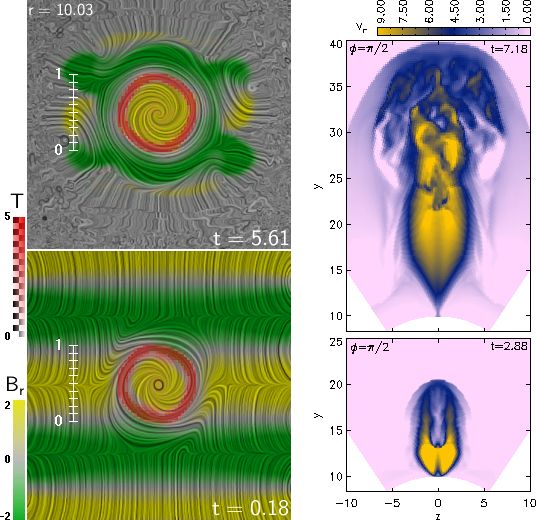}
\caption{{\it Left:} Magnetic field and temperature in a horizontal slice next
to the bottom boundary in a simulation with an arcade-shaped initial field
(setup D) with shear (Fig.~\ref{fig:emergecases}a) after 1 ({\it bottom left
panel}) and after 32 revolutions ({\it top left panel}).  The texture shows the
geometry of the parallel field component and the green/yellow coloring gives
the magnitude of the normal component.  Regions of increased temperature are
tinted in red (low temperatures are transparent).  {\it Right:} Evolution of
the radial velocity in a meridional slice.  The jet is eventually ``choked
off'' at the bottom.}
\label{fig:lic}
\end{figure}

Simulation setup D with the field geometry in Fig.~\ref{fig:emergecases}a was
found to produce extended, collimated outflows.  However, the magnetic field at
the base of the jet decays in this case (hence the ``D'') and is not
replenished, for which reason these jets are not permanent.

The forced rotation at the lower boundary stretches the magnetic field lines in
azimuthal ($\varphi$) direction around the axis of rotation ($y$).  The gas is
accelerated upwards and the magnetic field assumes a tangled helical structure.
A good momentary acceleration was obtained with the parameters
$(\betab=1/4,\Mb=18)$. The magnetic field in this case is depicted in
Fig.~\ref{fig:flinesD}.  The jet diameter is resolved with about 35 grid cells
in this simulation.  The jet attains a height of about 30 times its initial
diameter, the factor of expansion being about 7.  It reaches velocities that
are close to $\Mb$ and about a factor of $10$ above the escape speed.  The
field lines are closing with the lower boundary: inside the jet, the radial
magnetic field has the same polarity as on the disk, whereas the net radial
magnetic flux $\int B_r \, \de A$, integrated over the $r=\const$ surface, is
virtually zero at all times and all radii.

The ``absolute flux'' $\int \abs{B_r} \de A$ decreases linearly with time near
the lower boundary.  Increased temperatures near the outside of the rotating
surface indicate that a substantial amount of magnetic field is being
dissipated there, see Fig.~\ref{fig:lic}.  Due to the rotation, the magnetic
field assumes a vortex-like structure, at the border of which magnetic field
lines of opposite polarities become entangled.  This leads to a continuous
decay of the magnetic field.  Such a decay of the magnetic field, through
wrapping up of field lines followed by cancellation through diffusion, is
called ``convective expulsion'' in overturning flows
\citep[e.g.][]{1956Zeldovich,1963Parker}. The magnetic field in the physical
source of a successful jet evidently must be of a different nature. This calls
for a modification of the boundary conditions at the base of our simulations.
Such cases are discussed in the following sections.

\subsection{Jets from emerging fields: parameter study}

\begin{figure}[t]
\includegraphics[width=\linewidth]{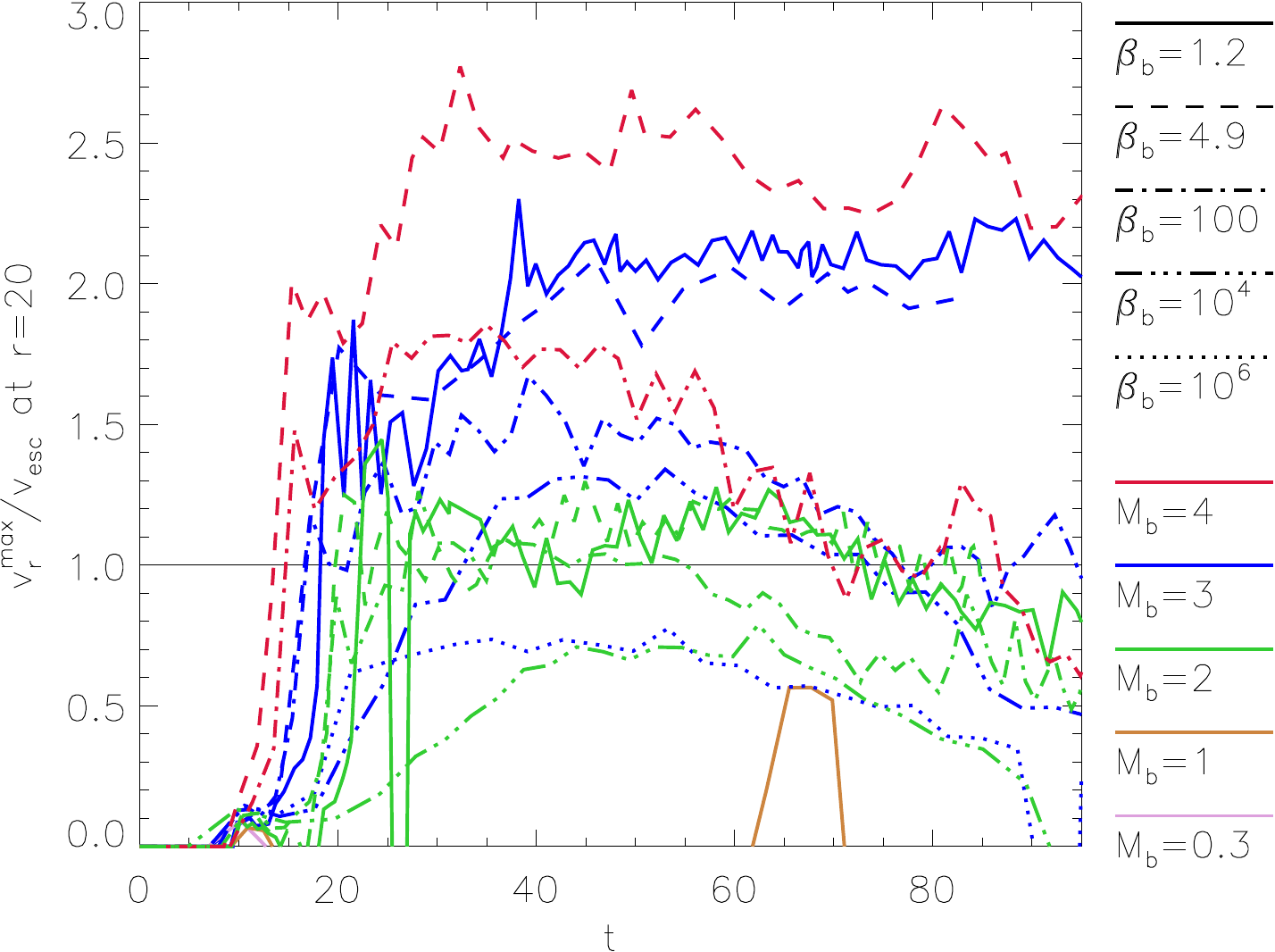} \\
\includegraphics[width=\linewidth]{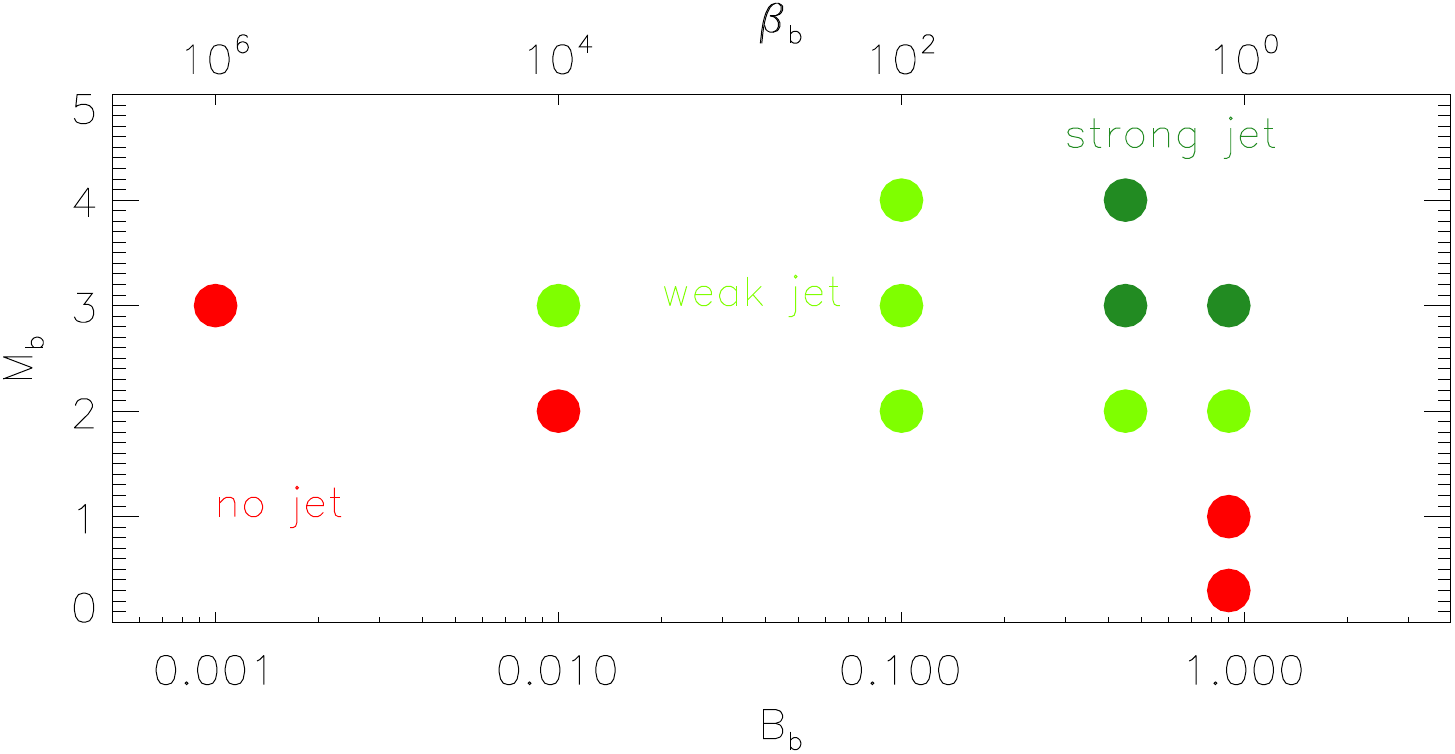}
\caption{\emph{Top panel}: Maximum velocity over escape speed for different
parameters $(\betab,\Mb)$ in otherwise similar simulations of the emerging
field model (E), variant (a).  The strength of the jet decreases with a
decrease of the field strength (increasing $\betab$, represented by different
line styles) and with a decrease of the rotation velocity ($\Mb$, represented
by different colors).  \emph{Bottom panel}: Strength of the jets as a function
of the rotation rate and field strength parameters.}
\label{fig:vrvesc}
\end{figure}

\begin{figure}[t]
\includegraphics[width=\linewidth,clip=true]{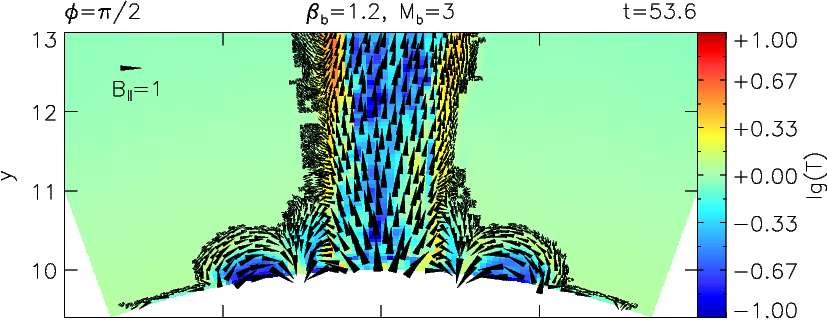} \\
\includegraphics[width=\linewidth,clip=true]{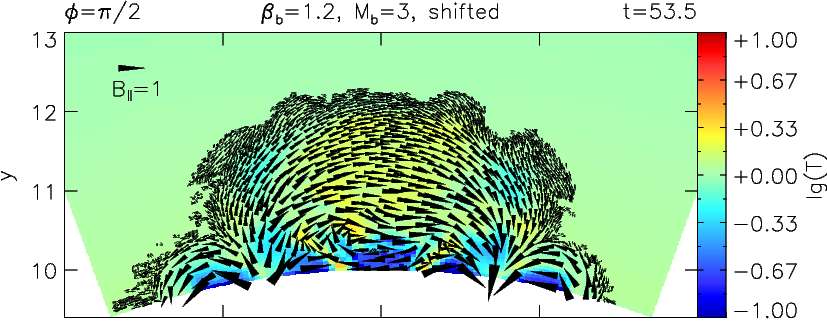} \\
\includegraphics[width=\linewidth]{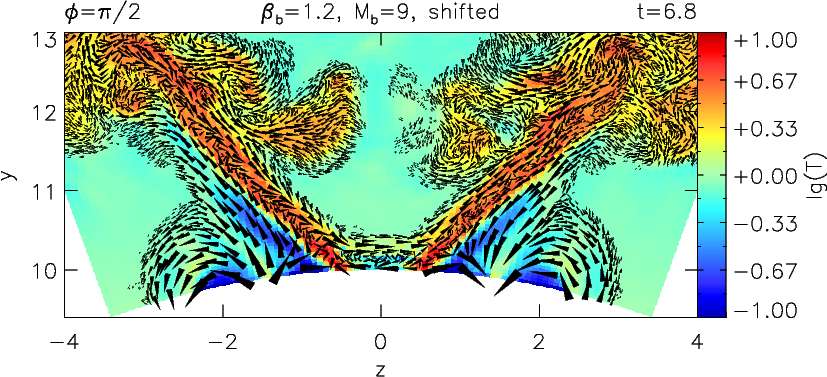}
\caption{Parallel magnetic field (wedges) and temperature (color) in a
meridional slice through the jet in a simulation with shear ({\it top panel})
and in two simulations without shear ({\it middle/bottom panel}). Collimated,
magnetically driven jets form only in cases where a toroidal field develops
through a shear of the magnetic field.}
\label{fig:wedges}
\end{figure}

Simulation setup E, in which the magnetic field loops above the rotating
surface are continuously replenished, produced long-lasting jets that propagate
considerable distances.  Before presenting big simulations with long jets in
the next two sections, we discuss the influence of the parameters
$(\betab,\Mb)$ on the results by comparing a series of smaller, computationally
cheaper simulations in which these parameters are varied. Although the
proximity of the upper boundary possibly influences the results (see discussion
of instabilities below), the small simulations give clear indications about
which parameters yield efficient jets.

The simulations cover the relatively short distance of $20$ length units ($40$
disk radii), with $10<r<30$ and $\theta,\phi=\pi/2\pm\pi/9$, the resolution
being $256\times96\times96$.  The atmospheric density differs by a factor $27$
between the lower and upper boundaries.  The simulations were pursued until
$t\approx100$. In the cases where a jet was successfully launched, this
suffices for about 2--3 passages through the computational volume. The
computational cost for this was 1--3 wallclock days on 36 processors with MPI
parallelization.

To test the flow's ability to penetrate the atmosphere in the setup with shear,
Fig.~\ref{fig:emergecases}a, we compare the jet velocities halfway through the
simulated distance against the escape velocity in Fig.~\ref{fig:vrvesc}.  The
jet velocity decreases with both decreasing rotation velocity $\Mb$ and
decreasing magnetic field strength $\Bb$ (or increasing $\betab$). It is,
however, much more sensitive to the former parameter: $\betab$ must be changed
by orders of magnitude to get a significant impact on jet velocity whereas with
$\Mb$, a factor of order unity suffices.  The bottom panel in
Fig.~\ref{fig:vrvesc} shows which flows pass this test, with those exceeding
the escape speed marked in green, the others in red.  Failed jets do not reach
the upper boundary.  The background atmosphere, whose equilibrium is perturbed,
tends to fall down on them.  The fixed conditions at the lower boundary avert a
pile-up of thermal energy: downward flowing hot gas vanishes across the
boundary and the injected gas has a constant temperature. A delayed onset of a
flow due to accumulated heat is unlikely for this reason.

There are instabilities in all cases where a jet develops.  The instabilities
develop mainly after the first passage through the computational domain, in the
form of helical displacements and/or a change of direction of the whole jet by
up to several degrees. The latter are likely modes with wavelengths longer than
the computational domain, i.e. one sees only the lower part of what would be a
kink if the radial extent of the domain was larger.  Jets created by stronger
magnetic fields ($\betab \le 4.9$) tend to show this kind of incipient
instability.  Jets from weaker fields move slower and develop pronounced
helical deformations already within the computational domain.  The proximity of
the upper boundary does not allow for more conclusive statements about
differences in instability behavior within the limits of this parameter study.
A series of large and expensive simulations of the kind presented in the next
section would be needed for that.

If the configuration is shifted such that no shear in the magnetic field occurs
(Fig.~\ref{fig:emergecases}b), a jet does not develop. The middle panel in
Fig.~\ref{fig:wedges} shows such a case.  The magnetic field lines are
unconnected to the surrounding atmosphere and the field is not amplified by
shear.  This appears to be sufficient to allow them to rotate without producing
a magnetically powered flow.  However, for very large values of $\Mb$ and
$\betab$, an outflow forms at the edge of the rotating disk, see bottom panel
in Fig.~\ref{fig:wedges}. This outflow appears to be driven by thermal buoyancy
associated with the dissipation of rotational energy at the disk's edge.
Having the shape of a hollow cone with a large opening angle, it is clearly
distinctive from the jets discussed above and the rest of this paper.

\subsection{Jets from emerging fields: a large simulation}
\label{sec:emlarge}

\begin{figure*}[t]
\includegraphics[width=\linewidth]{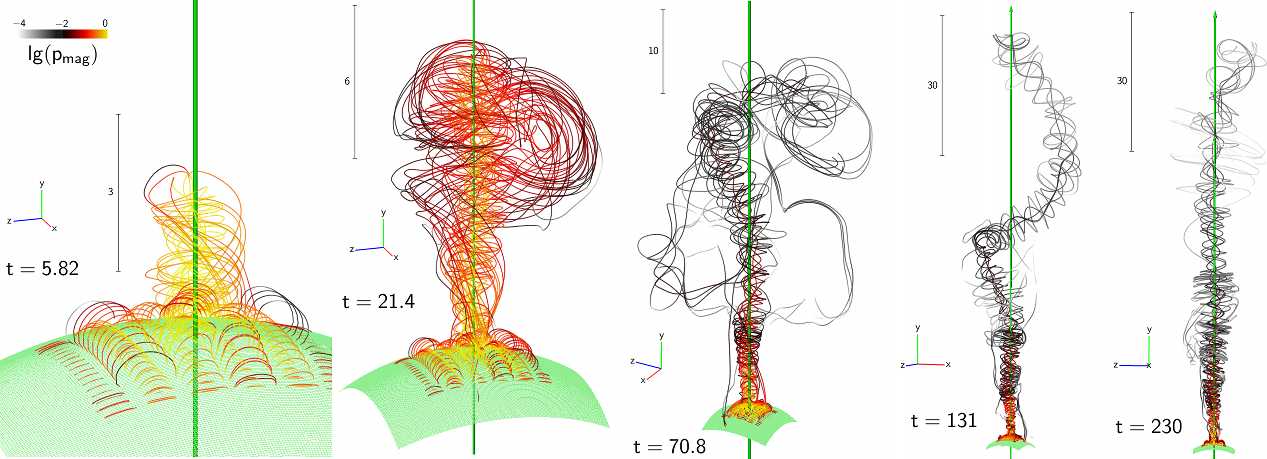}
\caption{Selected magnetic field lines in the simulation presented in
Sect.~\ref{sec:emlarge}.  The color gives the magnetic field energy. The
boundary conditions create magnetic field arcades above the boundary, the
arcades are twisted in the center by the imposed rotation and a jet with
helically shaped field lines forms along the axis of rotation.  The jet is
subject to occasional helical deformations.  Movies of this and other
simulations can be found at
\protect\url{http://www.mpa-garching.mpg.de/~rmo/pap3/index.html}.}
\label{fig:flinesE}
\end{figure*}

\linespread{0}
\begin{figure}[t]
\begin{tabular}{@{}l@{\hskip-.5pt}l}
\includegraphics[width=.52868181\linewidth,clip=true,hiresbb=true]{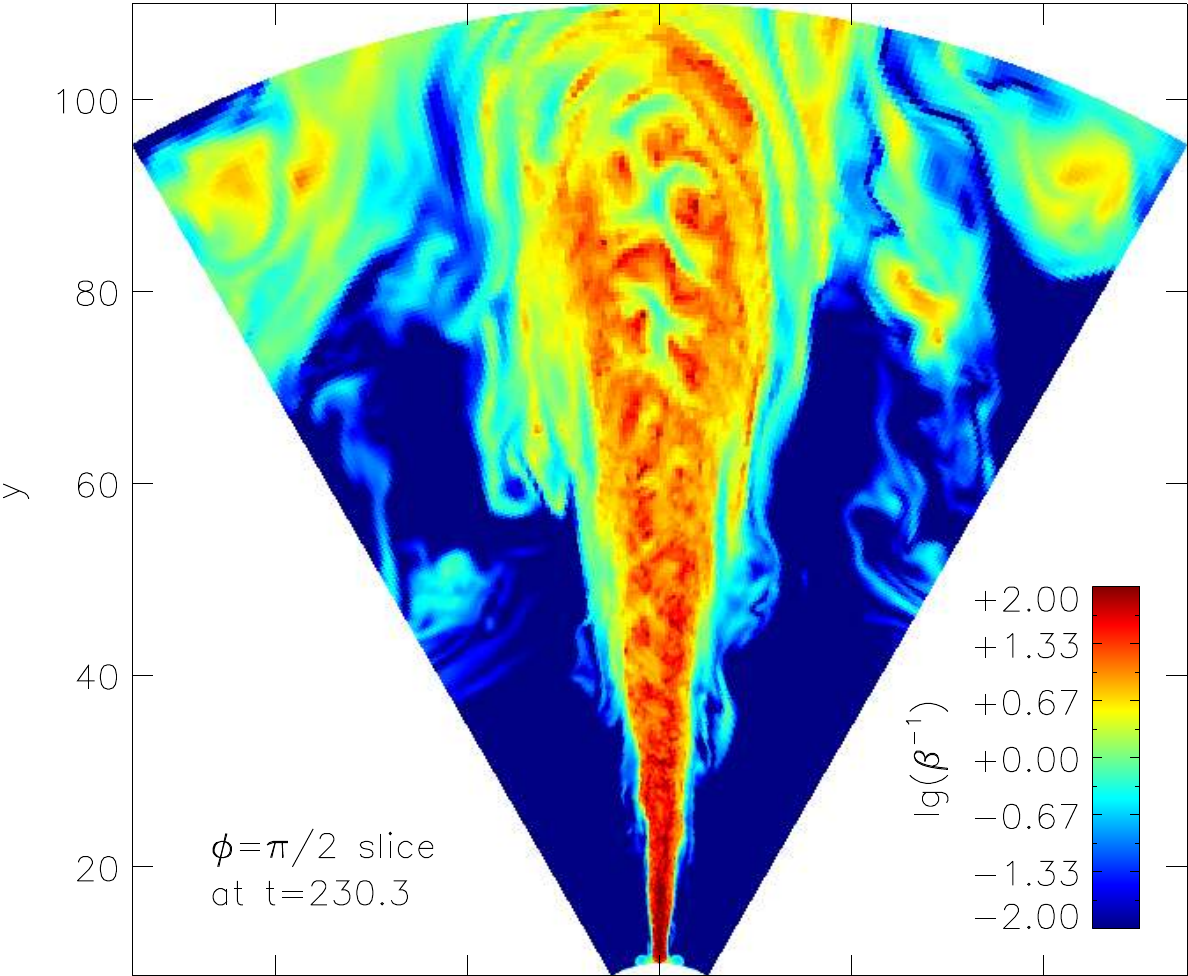} &
\includegraphics[width=.47131819\linewidth,clip=true,hiresbb=true]{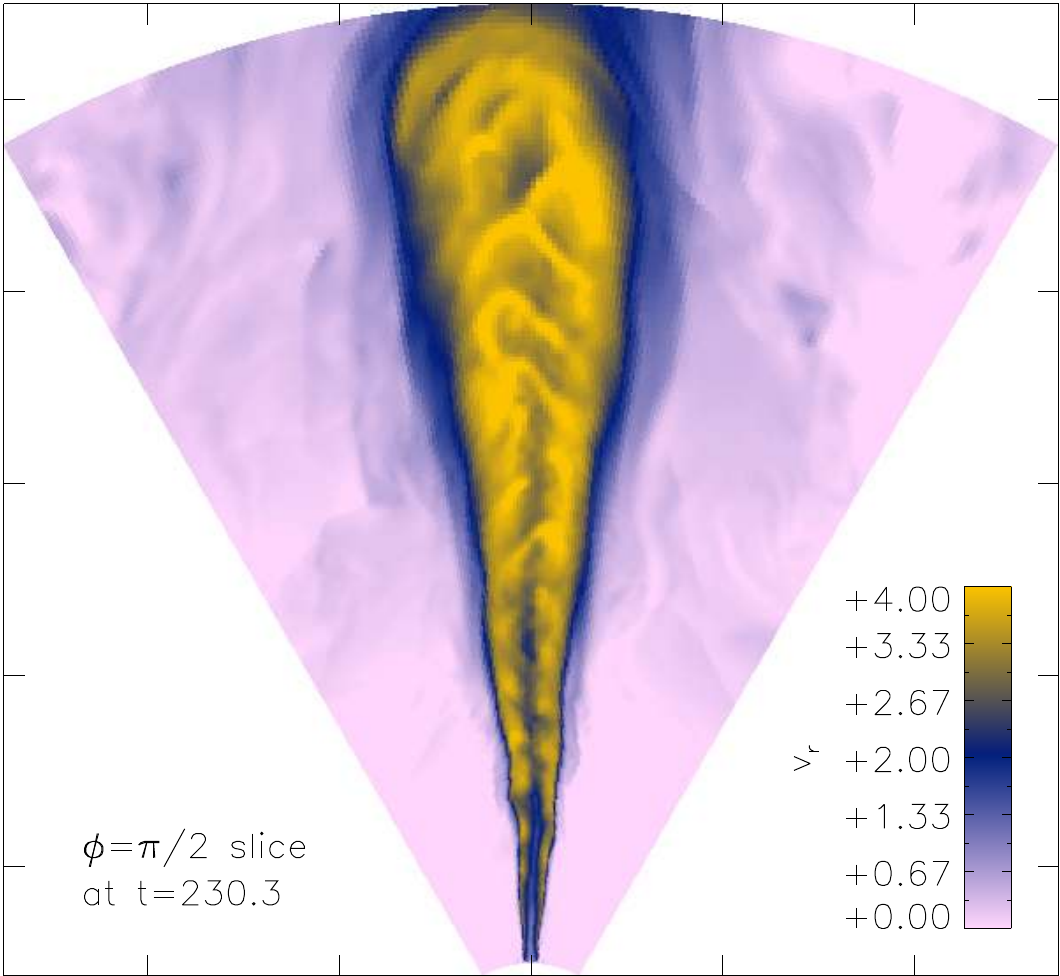}\\[-.5pt]
\includegraphics[width=.52868181\linewidth,clip=true,hiresbb=true]{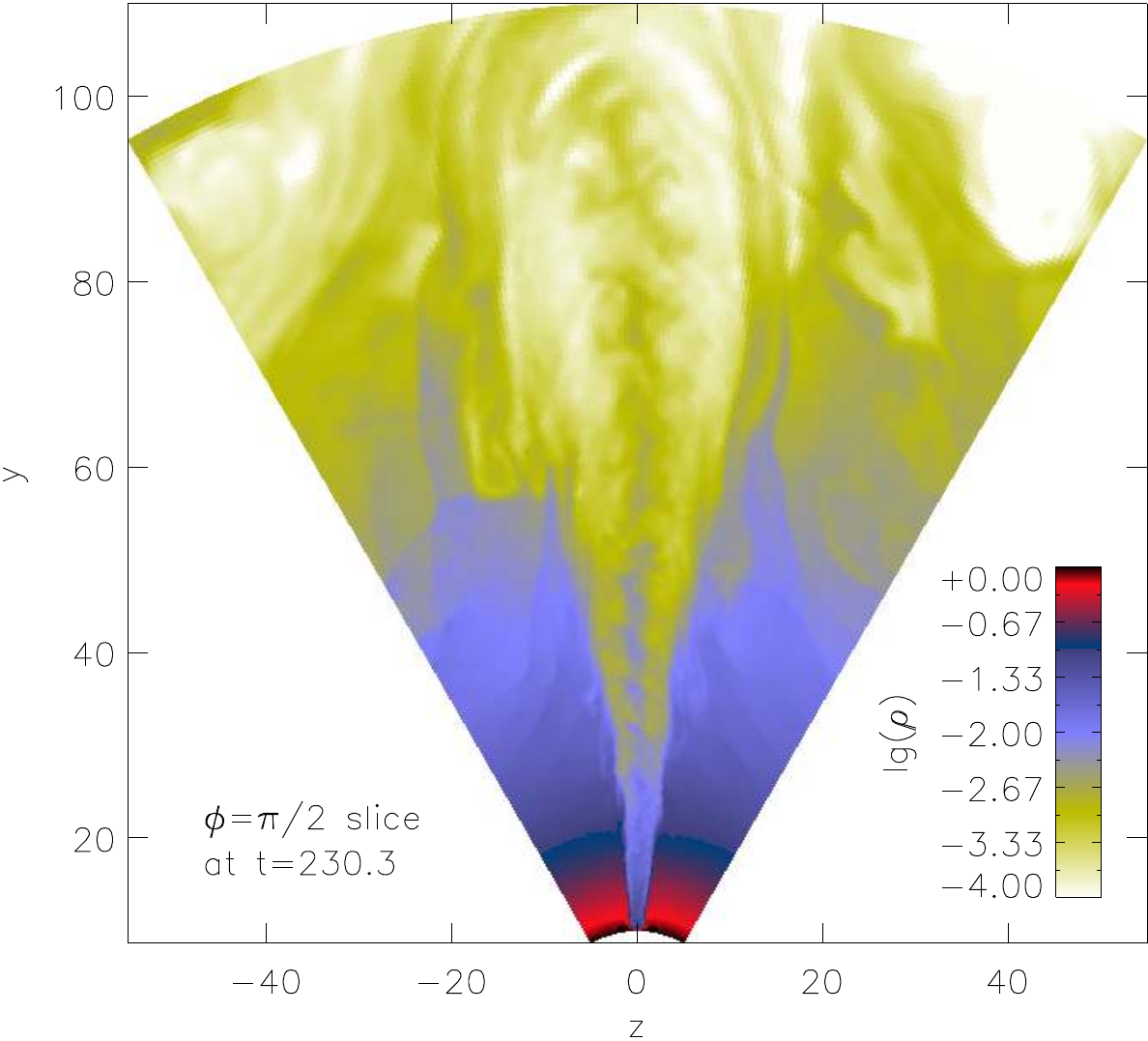} &
\includegraphics[width=.47131819\linewidth,clip=true,hiresbb=true]{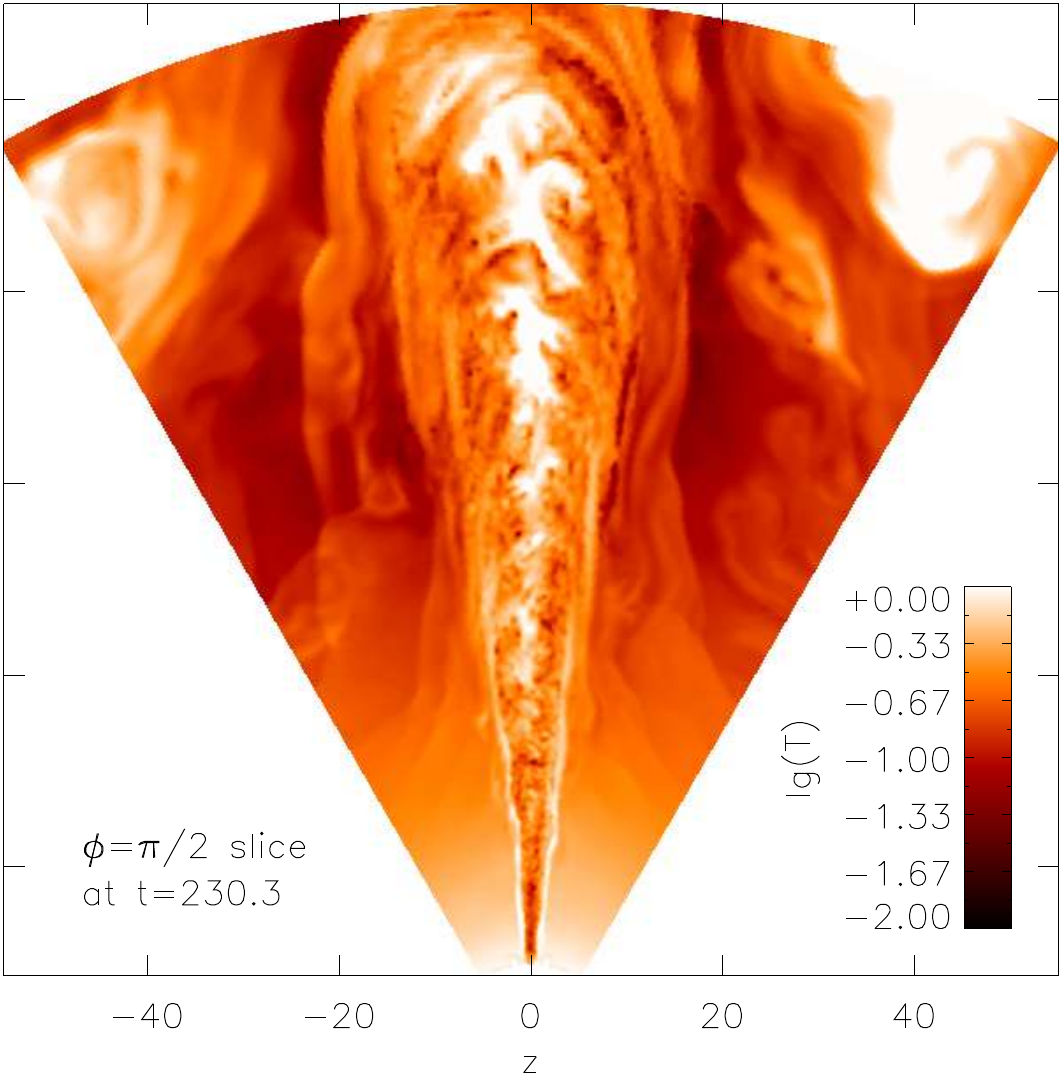}
\end{tabular}
\caption{Meridional slices through the jet presented in
Sect.~\ref{sec:emlarge}.  The magnetic jet is accelerated into a non-magnetic
atmosphere, penetrating through denser material.  The hot, magnetic material in
the environment of the upper part of the jet are remnants of previous
instabilities.}
\label{fig:slices}
\end{figure}
\linespread{1}

\begin{figure}[t]
\centering
\includegraphics[width=.9\linewidth]{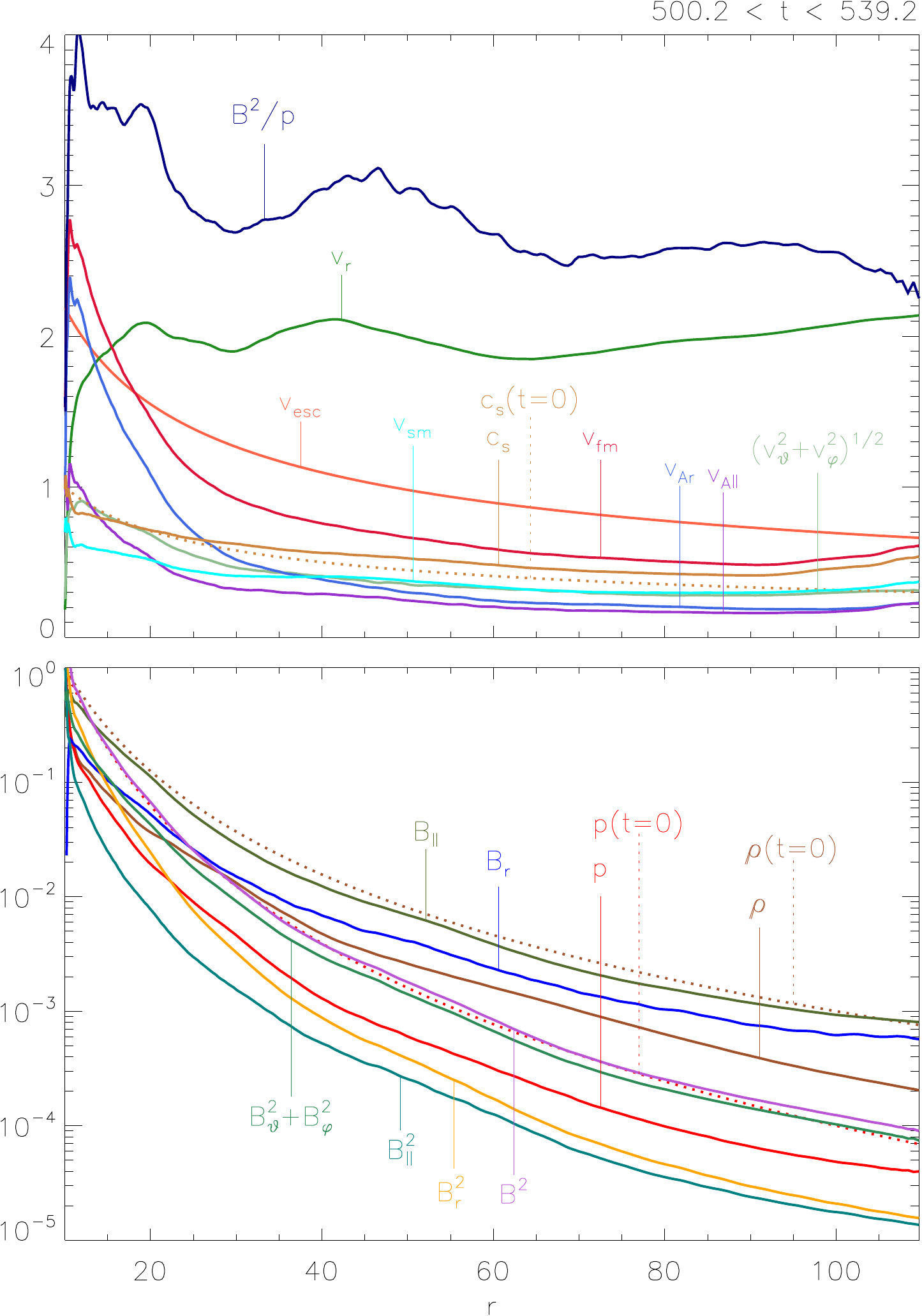}
\caption{Spatial and time-averaged values of various quantities in the
simulation presented in Sect.~\ref{sec:emlarge}.  The dotted lines represent
the initial atmospheric value, the $\parallel$ symbol stands for ``parallel to
$\vec{v}$''.}
\label{fig:u1meanvals}
\end{figure}

\begin{figure}[t]
\centering
\includegraphics[width=.9\linewidth]{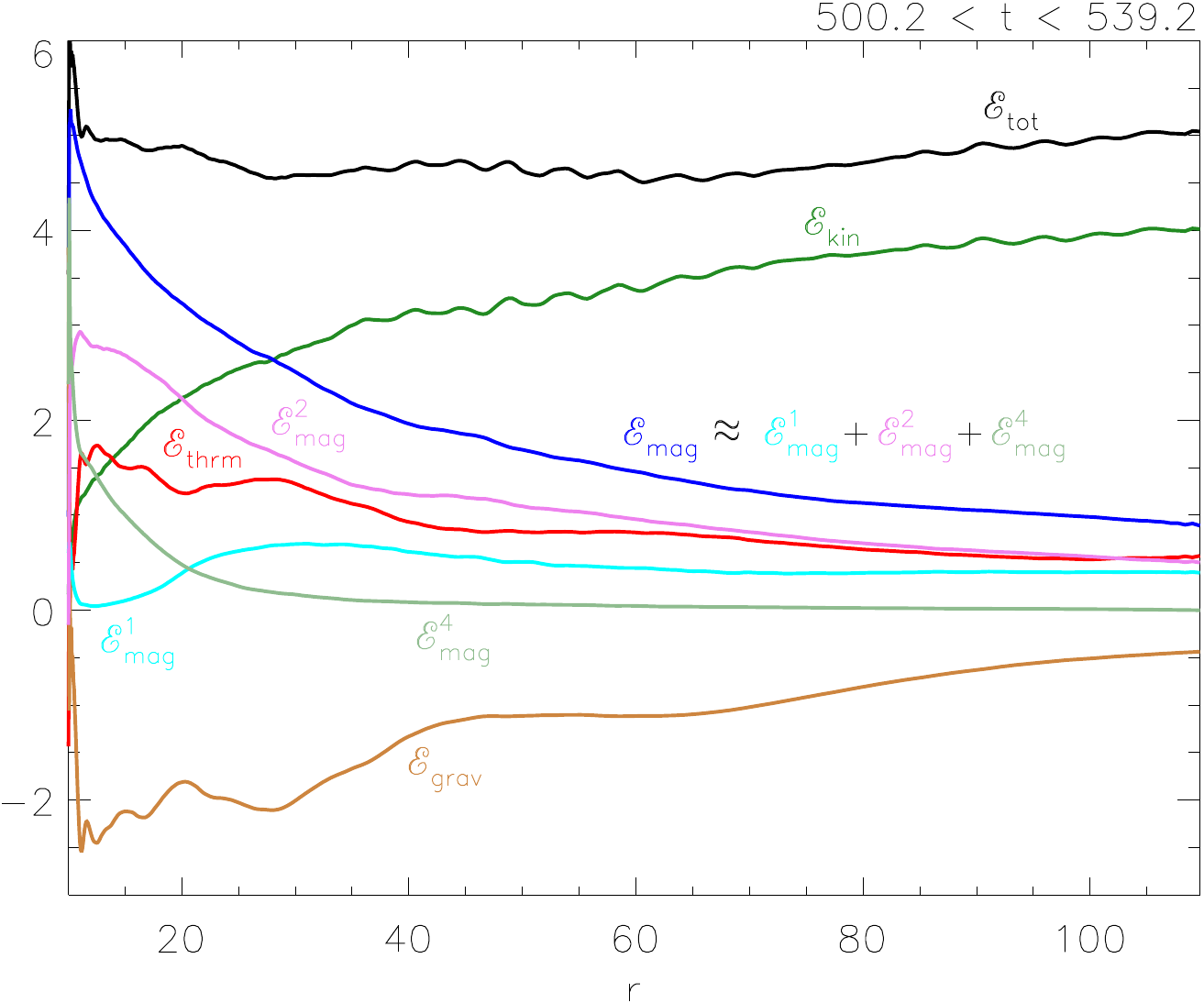}
\caption{Energy flow rates in the jet presented in Sect.~\ref{sec:emlarge}.
Magnetic enthalpy (integral over Poynting flux, blue line) is converted into
kinetic energy (green line) and potential energy (brown line).}
\label{fig:eflow}
\end{figure}

The long-range behavior of jets from emerging fields (setup E) in the case with
shear (Fig.~\ref{fig:emergecases}a) was investigated in a large simulation that
covers the radial range $10<r<110$. Along this range, the atmospheric density
decreases by a factor $1300$.  The parameters used were $(\betab=4,\Mb=3)$.
The jet crosses the upper boundary at the physical time $t\approx110$, which
corresponds to $105$ complete rotations of the disk; the simulation was stopped
at $t=539$.  For this, 19 wallclock days on 64 processors with MPI
parallelization were needed, the resolution being $400\times160\times160$.  The
lateral boundaries are at $\theta,\phi=\pi/2 \pm \pi/6$; the jet stays within
these boundaries at all times.

The wound-up magnetic field rises upwards as shown in Fig.~\ref{fig:flinesE},
with a collimated outflow forming in vertical direction.  Not counting the
changes in direction caused by instabilities, the jet's half opening angle is
about $5\degree$.  Inside the jet, the field lines spiral upwards, with the
radial magnetic flux having the same sign as the radial field on the rotating
surface.  The total flux is, in comparison, close to zero at most radii and
times. Accordingly, visualizations of the field lines give the appearance that
most lines remain connected to the lower boundary with both ends until the jet
crosses the upper boundary.

Fig.~\ref{fig:u1meanvals} shows the average value of various quantities across
the flow, i.e.  the average of a quantity $X$ is calculated as $\int X v_r\,\de
A / \int v_r\,\de A$ over the $r=\const$ surface.  The average velocity
increases mainly below $r \approx 20$, at which point it exceeds the average
fast magnetosonic speed $\vfm^2=\cs^2+\vA^2$.  The peak velocity (maximum of
$v_r$ on $r=\const$) is constant beyond $r \approx 30$ and is twice as large as
the average velocity ($4$ instead of $2$).  The average density is at all radii
significantly lower than the atmospheric value, the temperature ($\cs^2$) is
increased with respect to the environment.  The transversal (toroidal) magnetic
field dominates over the radial component at medium and large distances.  The
magnetic pressure dominates over the gas pressure by a factor of about 2--4 at
all distances in the jet.

The dependence of the energy flow rates with distance gives information about
energy transformations taking place in the jet. The total energy flow rate
$\efrate_\text{tot}(t,r)$ is obtained from a surface integral over the components of
the radial energy flux
\begin{equation}
    \underbrace{\frac{1}{2}\rho v^2 v_r}_\text{kinetic} + \underbrace{\frac{\gamma}{\gamma-1}p v_r}_\text{thermal enthalpy}
    + \underbrace{\rho \Phi v_r}_\text{grav. potential} + \underbrace{S_r}_\text{magnetic enthalpy} ,
\label{eq:eflux}
\end{equation}
with $S_r$ being the radial component of the Poynting vector:
\begin{equation}
    S_r = \frac{1}{4\pi} \left( B_\vartheta^2 v_r + B_\varphi^2 v_r
    - B_\vartheta B_r v_\vartheta - B_\varphi B_r v_\varphi \right) .
\label{eq:comps}
\end{equation}
We denote the individual components of $\efrate_\text{tot}$ by
$\efrate_\text{kin}$, $\efrate_\text{thrm}$, $\efrate_\text{grav}$ and
$\efrate_\text{mag}$ in order of their appearance in Eq.~\eqref{eq:eflux}. The
components of the magnetic enthalpy flow rate are denoted by
$\efrate_\text{mag}^{1\ldots4}$ in order of their appearance in
Eq.~\eqref{eq:comps}.  The energy flow rates in the present simulation are
plotted in Fig.~\ref{fig:eflow}.  The plot shows how magnetic enthalpy is
converted into kinetic and potential energy.  The conversion is efficient in
that less than $25\%$ of the initial $\efrate_\text{mag}$ remains in the jet
when it reaches the upper boundary.  Thermal energy is less important,
$\efrate_\text{thrm}$ being reduced only by half as many units as
$\efrate_\text{mag}$.  The most important components of $\efrate_\text{mag}$
are $\efrate_\text{mag}^2$, which is associated with the advection of the
azimuthal field and, at low radii, $\efrate_\text{mag}^4$, the work done by the
azimuthal flow against the azimuthal component of magnetic stress.
$\efrate_\text{mag}^3$ is virtually zero.  In summary, the qualitative behavior
of the energy flow rates is similar to those in simulations of jets generated
by twisting a uniformly polarized large-scale magnetic field (Papers I\&II).

The jet has a lot of substructure, see Fig.~\ref{fig:slices}.  Above $r \approx
20$, it is affected by recurrent instabilities that divert its course away from
the $y$-axis by several degrees.  Stirred up ambient material contributes to
the unstable behavior.  Occasionally, the jet develops a pronounced helical
shape as is characteristic for kink instabilities. The helix makes
approximately one complete turn within the computational volume, out of which
it is rapidly advected. For comparison, the magnetic pitch is
$\mathord{\lesssim} 10$ inside the jet (cf. Fig.~\ref{fig:flinesE}), which is
at least a factor 10 below the wavelength of the prominent instability. The
largest deflections from the central axis amount to approximately $10\degree$.
Being continually recreated at its base, the jet survives in a time-averaged
sense.

\subsection{Jets from a magnetic arcade on a differentially rotating surface}
\label{sec:singarc}

\begin{figure}[t]
\includegraphics[width=\linewidth]{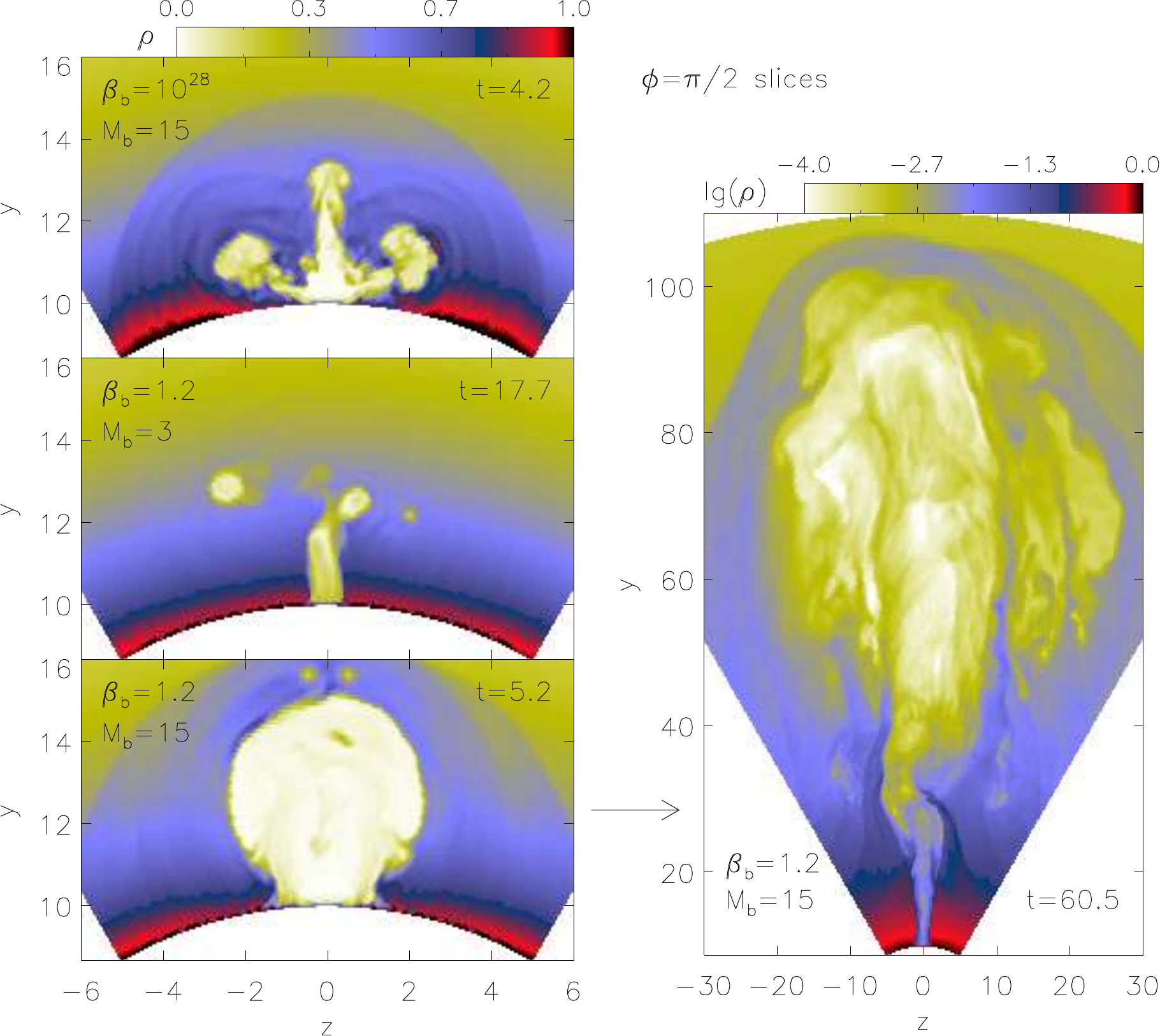}
\caption{Meridional slices through the jets presented in
Sect.~\ref{sec:singarc}.  A jet forms only if a magnetic field is present and
the rotation velocity is sufficiently large. Movies of such jets can be found
at the web address stated in Fig.~\ref{fig:flinesE}.}
\label{fig:vxrho}
\end{figure}

We also produced jets with emerging field loops (setup E) on a differentially
rotating surface, see sketch in Fig.~\ref{fig:emergecases}c.  These jets are
mostly the result of thermal buoyancy, driven by dissipative heating of the
near-disk atmosphere. Nonetheless, the presence of a magnetic field is required
to launch a directed flow.

The simulations cover the same physical range as the one presented in the
preceding section. The resolution is $400\times256\times256$, so that the
diameter of the differentially rotating surface is resolved with about 25
cells.  The longest simulation, which also produces the most efficient jet, has
the parameter values $(\betab=1.2,\Mb=15)$.  It ran for 14 wallclock days on 96
processors and covers almost two jet crossings, the upper boundary being
reached at $t\approx 62$.

In the jet of the preceding section, Poynting flux was the main source of
energy to accelerate the jet. This is clearly different here.  The simulation
with $(\betab=1.2,\Mb=15)$ attains a momentary quasi-stationary state, with
$\efrate_\text{tot}\approx\const$ in the energy flow, at $t \approx 80 \ldots
90$.  In sharp contrast to Fig.~\ref{fig:eflow}, $\efrate_\text{thrm}$ is the
dominating component at the beginning, being a factor of $\simm3$ larger than
$\efrate_\text{mag}$.  The equipartition point
$\efrate_\text{kin}=\efrate_\text{mag}$ is very close to the lower boundary,
the radius where $\efrate_\text{kin}=\efrate_\text{thrm}$ is larger, at
$r\approx15$.  $\efrate_\text{mag}$ also diminishes (by about 80\%), but it is
only a small fraction of the final $\efrate_\text{kin}$.

The excess temperature responsible for the high $\efrate_\text{thrm}$ turns up
near the central axis next to the lower boundary and increases with the speed
of rotation.  The heat is apparently produced by numerical viscosity, i.e.
dissipation of kinetic energy due to the coarse resolution of the rapidly
rotating inner part of the disk.  Unfortunately, a much higher resolution is
unfeasible and a large $\Mb$ is needed for an efficient production of
$B_\varphi$. The simulations are therefore relevant for cases in which
dissipative heating is of importance.

The appearance of the flow becomes more jetlike with increasing rotation
velocity.  In a simulation with $(\betab=1.2,\Mb=3)$, corresponding to a
differential rotation velocity of about $1.1$ of the outer loops' footpoints,
the jet terminates at a few disk radii from the source in a tightly wound
helix, similar to the free end of a garden hose.  With $\Mb=9$, the flow has a
turbulent, ``smoke stack''-like appearance.  The jet produced with $\Mb=15$ is
subject to non-axisymmetric perturbations, but much more coherent.

Despite the main energy source being thermal, the results depend greatly on
whether a magnetic field is present or not. In a simulation with
$(\betab=10^{28},\Mb=15)$, i.e. with virtually no magnetic field, there is no
unidirectional flow. Rather, material is ejected sideways as well as in the
upward direction, see Fig.~\ref{fig:vxrho}.  The simulation soon crashes,
presumably due to too much heat accumulating.  The same simulation with
$\betab=1.2$ is well collimated and plows effortlessly through the dense parts
of the atmosphere.

\section{Summary and discussion}
\label{sec:discussion}

We produced collimated outflows in numerical simulations by rotating the
footpoints of arcade-shaped magnetic loops.  The longest of these jets cross a
computational domain which is two orders of magnitude larger than the size of
the source.

Flow acceleration via dissipative as well as non-dissipative processes was
observed.  The latter relies on the presence of shear in the form of
differential rotation of the footpoints of individual field loops, which
results into the development of a toroidal magnetic field component.  The
resulting jet has a helical magnetic field structure with field lines running
back to the source outside the jet.

The case of uniformly rotating magnetic loops has also been studied.  It turns
out to be markedly less effective at producing outflows; in fact, no jetlike
outflows were observed for uniformly rotating arcades within the numerically
accessible parameter range.  The shear between the rotating loop and the
stationary atmosphere surrounding it also produces a toroidal field component,
but this appears to be much less effective than direct shearing of field lines
inside the rotating source.

We find that the continuation of the outflow beyond the initial transient
depends on the way the magnetic field is maintained at the lower boundary.
Differential rotation acting on a non-axisymmetric field quickly dissipates the
field through convective expulsion, with the result being that the jet is
``choked off'' at the base.  We have compensated for this by adding an inflow
of magnetic field at the base.  This would represent, for example, the
emergence of loops of magnetic field into the atmosphere of an accretion disk.

Apart from geometric requirements, the most important parameter for efficient
jets turned out to be the (differential) speed of rotation of the footpoints.
In the models calculated, the necessary speed is of the order of the escape
velocity from the disk. This is more than what is needed in simulations with a
long-scale poloidal initial field (e.g. those presented in Paper~I).
Consequently, the critical points (sonic, \Alfven, fast magnetosonic) and the
equipartition point between the magnetic and kinetic energy flow rates are
reached at smaller distances.  The conversion of magnetic enthalpy (and the
components thereof) to kinetic energy is similar apart from that, being fairly
efficient.

In the adiabatic calculations presented here, thermal energy from
hydrodynamic and magnetic dissipation stays in the flow. This is in contrast
with our earlier calculations (Papers~I\&II), where we assumed optically thin
environments in which much of the dissipated energy is lost by radiation. The
present models assume no energy loss and are most relevant for optically thick
conditions. In the stratified atmosphere of these models, heating contributes
to the flow by thermal buoyancy.  In some of the cases presented
(Sect.~\ref{sec:singarc}), this is the dominant driving mechanism.  Though a
wound-up magnetic field is also present in these buoyant plume flows, they
probably do not qualify as magnetically driven. The non-axisymmetric
``stirring'' by the rotating magnetic field at the base that drives them may,
however, well be relevant for the case of a (rapidly) rotating core inside a
dense stellar envelope. Even though its field configuration is not of the right
kind to produce a magnetically powered outflow, dissipation of rotational
energy by ``stirring'' may still have powerful effects on the envelope.

The magnetic field in the jets presented here is strongly wound, with the field
lines making many turns along the length of the jet, see Fig.~\ref{fig:flinesE}
for an example.  Current-driven instabilities, in particular kink modes, are
therefore to be expected and can indeed be found in the simulations.  The
turbulent wiggling of the jet increases the interaction with the ambient
medium, leading to an increased entrainment of material.  At large distances
(larger than those covered by the simulations), it is possible that the field
configuration is affected by instability-induced dissipation of magnetic energy,
presumably with dynamical consequences for the jet.  The effect
would likely be the same as in jets launched with large-scale magnetic fields
(see Paper~II for that).  Kelvin--Helmholtz instabilities at the interface
between the jet and the ambient medium are suppressed by the relatively coarse
resolution used in the simulations.

The final state of the jet, at large distances from the source, depends on the
impact of instabilities in the region beyond the computational volume of the
simulations presented here. It is reasonable to believe that the jet continues
as a ballistic flow when the instabilities cease to be effective (due to the
expansion of the jet and a possible decay of the toroidal field).  A
``disruption'' of the jet would require a strong interaction with the medium
into which it propagates; the atmosphere at large distances is probably too
thin for that.

Although the simulations presented here demonstrate possible forms of
rotation-induced acceleration, they probably do not represent very realistic
models of actual accretion disks.  One may wonder what happens in a more
realistic scenario with more complicated or chaotically arranged magnetic field
loops emerging from the disk.  Obviously, magnetic reconnection would likely
play a major role in such a scenario. Nonetheless, the surviving magnetic field
is stretched out in azimuthal direction by the rotation, and the free energy in
the toroidal field is transformed into an outflow.  Temporary fluctuations are
likely to occur if the supply of magnetic field loops is not continuous. The
coherent length of the jet will then be determined by the speed with which
reconnection processes destroy the initial field by which the flow is produced.
The jet may flare up anew when new loops emerge. This effect could be
responsible for the temporary fluctuations in gamma ray bursts.

Loops of magnetic field can also be found in the solar corona and their
evolution there is likely connected with explosive ejections of material (CMEs,
see e.g. \citealt{2006Forbes} for a review).  One general idea is that the
differential rotation as well as convective motions at the solar photosphere
produces shearing, and the energy in the sheared/twisted field is set free in
an explosive event \citep[e.g.][]{1989vanBallegooijen,2009Zuccarello}.  In
contrast to the jet models discussed here, the footpoints of a solar magnetic
arcade is unlikely to be twisted by a (pure) azimuthal motion. Non-azimuthal
shearing is not likely to produce an extended helical field (as in
Fig.~\ref{fig:flinesD}) and the plasma acceleration is hence different from
that in a magnetically driven jet.  Also, it seems that the energy in CMEs is
first accumulated and then released in a sudden event, much unlike the rather
smoothly rising jets presented here.

\begin{acknowledgements}
The author thanks H.~C.~Spruit for fruitful discussions and a careful reading
of the manuscript, and M.~Obergaulinger for providing his MHD code.
\end{acknowledgements}

\bibliography{ref}

\end{document}